# Comparative Evaluations of Visualization Onboarding Methods

Christina **Stoiber**[a], Conny **Walchshofer**[b], Margit **Pohl**[c], Benjamin **Potzmann**[c], Florian **Grassinger**[a], Holger **Stitz**[d], Marc **Streit**[b] and Wolfgang **Aigner**[a]

[a]*St. Poelten University of Applied Sciences, St. Poelten, Austria*
[b]*Johannes Kepler University Linz, Linz, Austria*
[d]*datavisyn GmbH, Linz, Austria*
[c]*TU Wien, Vienna, Austria*

## ABSTRACT

Comprehending and exploring large and complex data is becoming increasingly important for a diverse population of users in a wide range of application domains. Visualization has proven to be well-suited in supporting this endeavor by tapping into the power of human visual perception. However, non-experts in the field of visual data analysis often have problems with correctly reading and interpreting information from visualization idioms that are new to them. To support novices in learning how to use new digital technologies, the concept of onboarding has been successfully applied in other fields and first approaches also exist in the visualization domain. However, empirical evidence on the effectiveness of such approaches is scarce. Therefore, we conducted three studies with Amazon Mechanical Turk (MTurk) workers and students investigating visualization onboarding at different levels: (1) Firstly, we explored the effect of visualization onboarding, using an interactive step-by-step guide, on user performance for four increasingly complex visualization techniques with time-oriented data: a bar chart, a horizon graph, a change matrix, and a parallel coordinates plot. We performed a between-subject experiment with 596 participants in total. The results showed that there are no significant differences between the answer correctness of the questions with and without onboarding. Particularly, participants commented that for highly familiar visualization types no onboarding is needed. However, for the most unfamiliar visualization type — the parallel coordinates plot — performance improvement can be observed with onboarding. (2) Thus, we performed a second study with MTurk workers and the parallel coordinates plot to assess if there is a difference in user performances on different visualization onboarding types: step-by-step, scrollytelling tutorial, and video tutorial. The study revealed that the video tutorial was ranked as the most positive on average, based on a sentiment analysis, followed by the scrollytelling tutorial and the interactive step-by-step guide. (3) As videos are a traditional method to support users, we decided to use the scrollytelling approach as a less prevalent way and explore it in more detail. Therefore, for our third study, we gathered data towards users' experience in using the in-situ scrollytelling for the VA tool Netflower. The results of the evaluation with students showed that they preferred scrollytelling over the tutorial integrated in the Netflower landing page. Moreover, for all three studies we explored the effect of task difficulty. In summary, the in-situ scrollytelling approach works well for integrating onboarding in a visualization tool. Additionally, a video tutorial can help to introduce interaction techniques of a visualization.

## 1. Introduction

Visualization can be seen as a process that transforms data into a visual form [14, 57]. As a user, this transformation needs to be traceable to decode the visual representation and correctly reason about the data. Albeit humans are visual beings and visual representations are easier to understand than other forms of data representations, we have to learn how to read and comprehend them. Unlike reading and writing a text, we are typically not taught how to read or interpret visualizations in the course of our education—with the exception of simple business charts, like bar, line, or pie charts, which usually encounter at a young age [2]. Hence, many users have difficulties interpreting and working with novel visual representations they are not familiar with [27, 50]. This not only bears the risk of drawing wrong conclusions but also leads to frustration or rejection of otherwise powerful data visualizations [10]. Boy et al. [11] describe visualization literacy as

"the ability to use well-established data visualizations (e.g., line graphs) to handle information in an effective, efficient, and confident manner". Having limited visualization literacy skills can be a serious handicap as it hinders people from valuable information retrieval which could be used to learn and solve problems, or make informed decisions [24, 11, 10].

Visual mapping is the process of assigning data variables to visual channels, which results in either a static or an interactive visual representation. This process is the central component of virtually all known conceptual models of visualization, such as the model by Card et al. [14] or van Wijk [70]. Understanding the process of visual mapping is a key component for correctly decoding both the visual representation and the underlying data. Furthermore, data analysis, filtering, and rendering steps of the visualization process influence the appearance of a visualization idiom and need to be consciously used and selected. However, especially for novice users, this can be difficult and may lead to wrong conclusions as well as erroneous insights into the data. Visualization onboarding may alleviate this and empower users to better understand the data and take full advantage of the visualization.

We define visualization onboarding [63, 65] as follows: *Visualization onboarding is the process of supporting users in reading, interpreting, and extracting information from visual representations of data.* A few onboarding methods exist in the literature using different strategies and educational theories, such as learning by doing [36], learning by analogy [56], scaffolding [7], or top-down and bottom-up teaching methods as well as active and passive learning types [67]. All of these concepts are stand-alone learning environments and are not totally integrated into a visual analytics (VA) tool itself. Only Yalçin [71] developed an in-situ help for the visualization tool *Keshif* [72]. Nonetheless, further research is needed to identify effective designs of onboarding methods and to understand users' behavior while using onboarding methods.

To fill this research gap, the main objectives of this research is to achieve an understanding on how visualization onboarding affects the user performance (see Figure 1). Therefore, we conducted two large-scaled studies with 388 and 145 MTurk workers (study 1—see Section 4 and study 2—see Section 5) and one study with 63 students (study 3—see Section 6). The qualitative results of study 2 have already been published in the short paper by [65]. We investigated various aspects of visualization onboarding (see Section 3), which we describe here in detail:

1. The effect of visualization onboarding using an interactive step-by-step guide by assessing the users' performance for four different interactive visualization types with varying complexity—a bar chart , a horizon graph  a change matrix , and a parallel coordinates plot , described in Section 4;

2. The effect on user performance for four different types of visualization onboarding methods. (i) a step-by-step guide, (ii) scrollytelling, (iii) a video tutorial, and (iv) an in-situ scrollytelling, see Section 5;

3. Differences in user performance based on three task types according to Friel et al. [23]: *reading the data*, *reading between the data*, and *reading beyond the data*;

4. Differences in the subjective user experience and answer correctness between an in-situ visualization onboarding concept using scrollytelling and a tutorial with videos in the VA tool *Netflower* [64] for a Sankey diagram  and a bar chart .

5. We present guidelines for the design of visualization onboarding methods, described in Section 7.3.

The overall aim of the studies is to provide a comprehensive overview of onboarding approach advantages using different types of visualizations and different types of onboarding methods (see Figure 1). We, therefore, analyze quantitative data on the users' performance and qualitative data such as textual feedback or subjectively rated attitudes and preferences from participants. The analysis of different visualization onboarding methods showed that tutorials and videos are common ways [63] to support users. Hence, we started by exploring a traditional method — an interactive step-by-step guide (tutorial) — using textual descriptions and visual markers inspired by the metaphor of legends for four different interactive visualization types with varying complexity. Results of study 1 revealed that onboarding is needed for more complex visualization techniques such as the parallel coordinates plots. We therefore further examined different onboarding methods. As a traditional approach besides the step-by-step guide, we developed a video tutorial. Besides, we wanted to investigate scrollytelling in the context of visualization onboarding. As a promising onboarding approach resulting from study 2, we further examined the applicability of a scrollytelling onboarding by embedding it into a VA tool (see study 3).

## 2. Related Work

So far, there has been little discussion about onboarding concepts for visualization techniques and VA tools [21]. The educational community started by studying how students interpret and generate data visualizations [5]. They investigated how to teach bar charts in early grades [2] using a tablet app, called *C'est la vis*, supporting elementary school pupils to learn how to interpret bar charts based on the *concreteness fading* approach. Concreteness fading is a pedagogical method where concrete examples are provided for abstract ideas and principles at first, before progressively abstracting them. Recently, Bishop et al. [7] developed a tablet-based tool called *Construct-A-Vis*, which supports elementary school children in creating visualization based on free-form activities. They used *scaffolding* as a pedagogical method which immediately provides feedback to the users if the visual mapping was correct. More recently, Firat et al. [20] developed an interactive pedagogical treemap application for training. The conducted study revealed that students who interacted with the tool outperformed students who only learned through slides before taking the literacy test. In this context, Echeverria et al. [18] performed first steps towards defining Data Storytelling to support teacher's sensemaking. The authors found out that the included narratives were helpful to support the story and to understand data points in the visualization. Further research has explored how data storytelling concepts can be used for communicating scientific data [41], presenting *data stories* to a broader audience [61], or supporting presenters to tell a story through data visualizations effectively [34].

Tanahashi et al. [67] investigated top-down and bottom-up teaching methods as well as active or passive learning types. The bottom-up teaching method ("textbook approach") [73] focuses on small, detailed pieces of information which students then combine to get a better understanding. Besides, a top-down teaching method is given when a broad overview first helps to understand the abstract, high-level parts of an idea/topic which then provide context for understanding its components in detail [67]. Furthermore, a distinction can be made between active and passive learning types. Passive learning means that students only receive the information without participatory dialog. In contrast, active learning describes an active participation [67]. Their analysis indicates that top-down exercises were more effective than bottom-up and active learning types with top-down tasks the most effective ones. In a comparative study by Kwon and Lee [36], the effectiveness of active learning strategies was ascertained. Three tutorial types—static, video-based, and interactive—were used to support the learning of scatterplot visualizations. Their observations show that participants using the interactive and video tutorials outperformed participants using a static or no tutorial at all. Ruchikachorn and Mueller [56] explored the learning-by-analog concept by demonstrating an unfamiliar visualization method by linking it to another more familiar one. The authors found out that the learning by analogy concept is useful as participants in their study could understand the unfamiliar visualization methods fully or at least significantly better after they observed or interacted with the transitions from the familiar counterpart. They assessed four combinations and compared their difference in visual literacy: scatter plot matrix against hyperbox, linear chart against spiral chart, hierarchical pie chart against treemap, and data table against parallel coordinates plots. The authors describe an additional advantage of learning-by-analogy over other forms of demonstrations such as textual or oral descriptions as they bridge any language barriers.

Besides scientific literature, onboarding concepts are integrated in commercial visualization tools as well. Nowadays most of these commercial visualization tools already integrate onboarding concepts focusing on the explanation of features. IBM Cognos Analytics [32], for example, uses step-by-step tours with tooltips and overlays for onboarding new users. A more traditional approach is used by the commercial visualization tool Advizor [1] which makes use of textual descriptions to explain the visual mapping for visualization techniques. Besides, there are platforms and websites available which can be categorized as external onboarding methods [63] supporting users in understanding the visual mapping of various visualization techniques. For instance, *The graphic continuum* [66] provides an overview of visualization types and supports design and method decisions. Similarly, the *Data Visualisation Catalogue* [54] seeks to support users to understand the encoding and building blocks of different visualization types. Furthermore, *From Data to Viz* [29] aims to find an appropriate visualization type based on the input data using a decision tree. The catalogues offers definitions, variations, and the use of each visualization type in addition to potential issues that may arise during use and interpretation. These systems are neither related to a particular visualization tool, nor do they integrate any educational theories. In recent literature, Wang et al. [69] present a set of cheat sheets to support visualization literacy around visualization techniques inspired by infographics and data comics, which are well-established onboarding methods in domains such as machine learning.

## 3. Research Design

To understand how visualization onboarding affects the user performance at different levels, we conducted three user studies. As a first step, we conducted a small-scaled preliminary study to understand how people step-wise explain the *WHAT*, *WHY*, and *HOW* of visualization techniques. The aim of this study was to use the findings as a design basis for the first interactive step-by-step onboarding method.

*Preliminary study:* Previous studies [31, 30] examined how people create, update, and explain their own visualizations using only tangible building blocks. Their main goal was to investigate how people construct their own visualizations using physical tokens — wooden tiles taken from a learning toys kit designed by Froebel [43] for Kindergarten education. To our knowledge, there are no studies available exploring how persons explain a certain visualization type. To provide a first step towards a deeper understanding, we conducted pilot interviews with 13 participants (m = 8, f = 5; Age: $M = 30.23$, $SD = 3.32$). The participants indicated that they had moderate to high experience with visualizations and a background in computer science, accounting, and HCI. For the interviews, we developed three different static visualizations— bar chart, horizon graph, and a change matrix. The bar chart showed sun hours in Innsbruck, Austria from 1965 to 2019. For the horizon graph we used random quantitative values from 0 to 30. Additionally, the change matrix visualized quantitative values ranging from -10 to 10. We asked the participants to explain the three visualizations. We took hand-written notes during the interviews. The systematic analysis showed that participants started explaining the data set and attributes first before continuing with the visual encoding, for instance for a bar chart: "The first thing I would say is that the diagram explains how many hours of sunshine the city of Innsbruck had in previous years." (translated from German to English). We have grouped the comments and explanations, and therefore we gained the following splitting of onboarding instructions: namely *reading*, *using*, and *interacting* with the chart, as described in Section 3.1.

In the following, we present the research questions we want to answer and elaborate on the different onboarding designs and the general study setup as illustrated in Figure 1.

*User studies 1–3: Study 1* was conducted with MTurk workers using LimeSurvey between 10/2019–11/2019 and 02/2020 and aimed to understand (1) if and how visualization onboarding affects the user performance with an interactive step-by step guide; (2) if and how the user performance between four different visualization types (bar chart, horizon graph, change matrix, and parallel coordinates plot) varies; and (3) if and what differences in the user performance can be observed for different task types [23].

We conducted the second study (*study 2*) from 08/2020 to 09/2020. The objective was to investigate how different visualization onboarding methods (interactive step-by-step guide, scrollytelling, or video tutorial) affect the user performance. Therefore, we developed and designed two further onboarding methods: a scrollytelling (in-situ onboarding) and a video tutorial. In line with study 1, we examined if a difference in user performance based on different task types exists. For both studies, we used a between-subject design, where each participant evaluated one out of four visualization types/onboarding methods, either with or without onboarding (independent variables). The level of experience for each visualization type was captured at the beginning of the survey. To identify an effect of onboarding, we posed questions directly before and after the interaction with the onboarding part. The questions include tasks with different difficulties [23]. We used different data sets for the questions before and after the onboarding part. We refer to the question set before the onboarding part as *pre-tasks*, and to the questions after the onboarding as *post-tasks*. To compare the performance measures for study 1, we coded the answer correctness with "0" when the question was incorrectly answered, with "1" when the answer was correctly given. Similarly, with respect to the improvement measures, we coded answers that were correctly answered in the pre-question set and incorrectly answered in the post-question set with "0", answers that were initially incorrectly answered and then in the post-question set correctly answered with "1", and questions that were equally correct or equally incorrect answered with "-99" as no improvement or deterioration was achieved. Similar for response time improvements, we evaluated the time difference between pre- and post-tasks. Thus, the more negative the value is, the faster the post-task was answered in contrast to the pre-task counterpart.

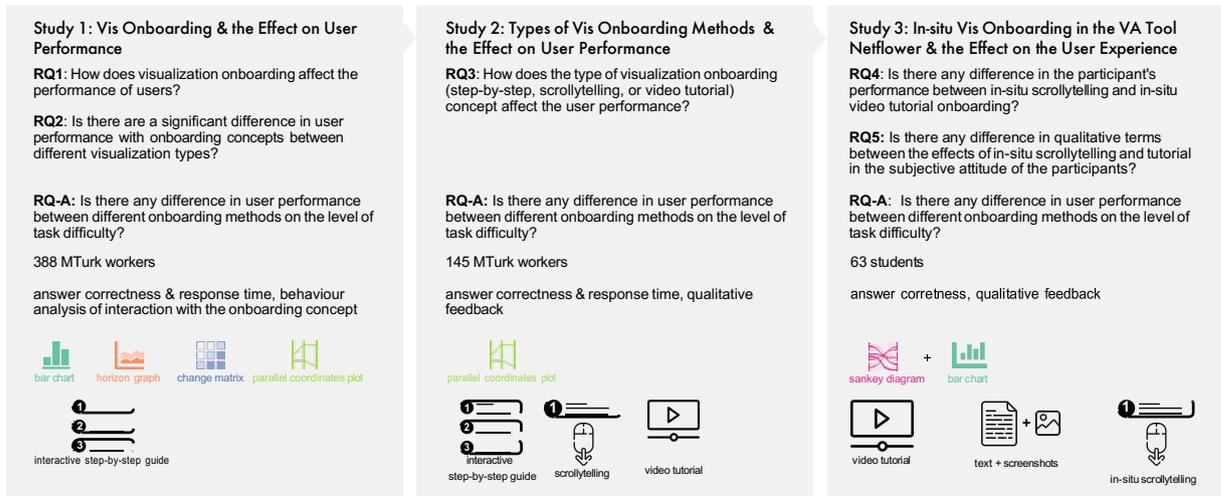

Figure 1: Overview of the three studies that were performed by highlighting the respective research questions (RQ), the origin of the participants as well as the performance metrics, visualization types used, and onboarding technique applied. We started with exploring a traditional method — an interactive step-by-step guide (tutorial) — using textual descriptions and visual markers inspired by the metaphor of legends for four different interactive visualization types with varying complexity. Results of study 1 revealed that onboarding is needed for more complex visualization techniques such as the parallel coordinates plots. We therefore further examined different onboarding methods. As a traditional approach besides the step-by-step guide, we developed a video tutorial. Besides, we wanted to investigate scrollytelling in the context of visualization onboarding. As a promising onboarding approach resulting from study 2, we further examined the applicability of a scrollytelling onboarding by embedding it into a VA tool (see study 3).

Lastly, from 06/2020 to 07/2020 and 10/2020 to 11/2020,

we conducted the third study (study 3) to understand how in-situ visualization onboarding in the VA tool *Netflower* (Sankey diagrams and bar charts [64]) affects both the user performance and the user experience. For this study, we aim to enable comparative evaluation with students between the provided onboarding in the Netflower tool and the in-situ method.

To elaborate on the effect of onboarding methods and designs as outlined in Figure 1, we present five research questions (**RQ**):

**RQ1** How does visualization onboarding affect the performance of users?

**RQ2** Is there a significant difference in user performance with onboarding concepts between different visualization types?

**RQ3** How does the type of visualization onboarding (step-by-step, scrollytelling, or video tutorial) concept affect the user performance?

**RQ4** Is there any difference in the participant's performance between in-situ scrollytelling and in-situ video tutorial onboarding?

**RQ5** Is there any difference in qualitative terms between the effects of in-situ scrollytelling and tutorial in the subjective attitude of the participants?

Moreover, we investigated the following research question for all three studies:

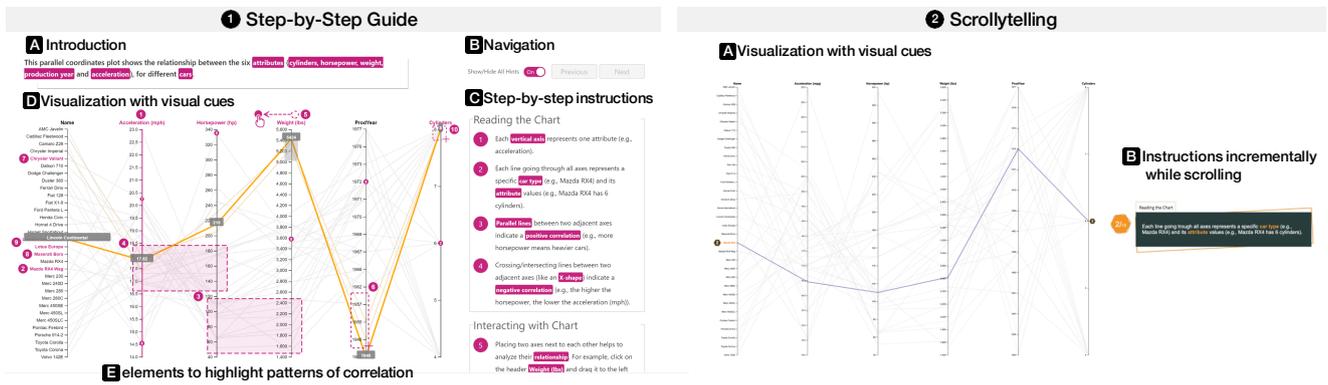

Figure 2: ❶ **Step-by-step guide for a Parallel Coordinates Plot.** The step-by-step guide is based on textual descriptions and in-place annotations and consist of four parts: a brief textual introduction providing contextual information about the visualization Ⓐ, navigational elements Ⓑ to go through the step-by-step instructions Ⓒ, and the visualization itself Ⓓ. We divide the textual descriptions into *Reading the Chart*, for explaining the visual encoding, *Interacting with the Chart*, for explaining the interaction concept, and *Using the Chart*, for providing exemplary insights. ❷ **Scrollytelling tutorial.** Users can incrementally scroll through the instructions Ⓑ while the visualization Ⓐ changes to the text appropriately.

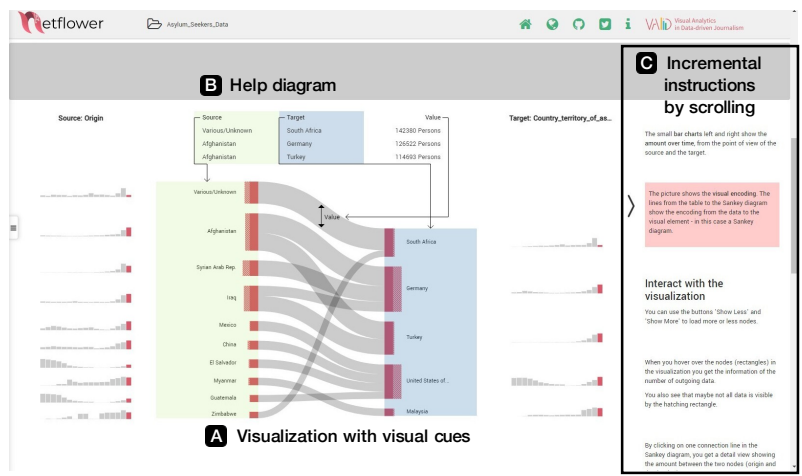

Figure 3: **In-situ onboarding for *Netflower*.** The main visualization Ⓐ and the onboarding panel Ⓒ on the side. The user can incrementally scroll through the onboarding messages. All supplementary components are hidden and only revealed in tandem with their explanation. To combat change blindness, the current onboarding message as well as the explained elements are highlighted in red Ⓒ. Additional help diagrams Ⓑ are overlaid when appropriate, using the actual data points visualized.

**RQ-A** Is there any difference in user performance between different onboarding methods on the level of task difficulty?

### 3.1. Onboarding Methods

In this section, we provide the description of the four different visualization onboarding methods and it's design decisions: step-by-step guide, scrollytelling tutorial, video tutorial, and in-situ onboarding using Netflower.

*Step-by-step Guide:* The design of our onboarding concept for the *step-by-step guide* is inspired by interactive legends as known from maps (e.g., thematic maps and cartographic communication) and statistical charts. Printed maps traditionally use static legends to encode the meaning of symbols and colors used [26]. In general, legends are an essential design element for cases where data can no longer be labeled directly in visualizations [19]. We took up the metaphor of legends and developed an interactive visualization onboarding concept consisting of step-by-step textual descriptions for both the visual encoding and interaction techniques (e.g., hovering, selecting, etc.) and combine them with in-place annotations in the main visualization. As described in Section 3, we divided the onboarding instructions into three parts: *reading*, *interacting*, and *using* the chart, illustrated in Figure 2 (left). In the next paragraphs, we elaborate on the different parts. **Introduction** Ⓐ: Due to the importance of title elements and legends [8], we integrated an introductory sentence at the top of our onboarding method to provide general information about the data and describe the area of application for the respective visualization type. For the parallel coordinates plot, for instance, the purpose of the visualization is to identify relationships between attributes.

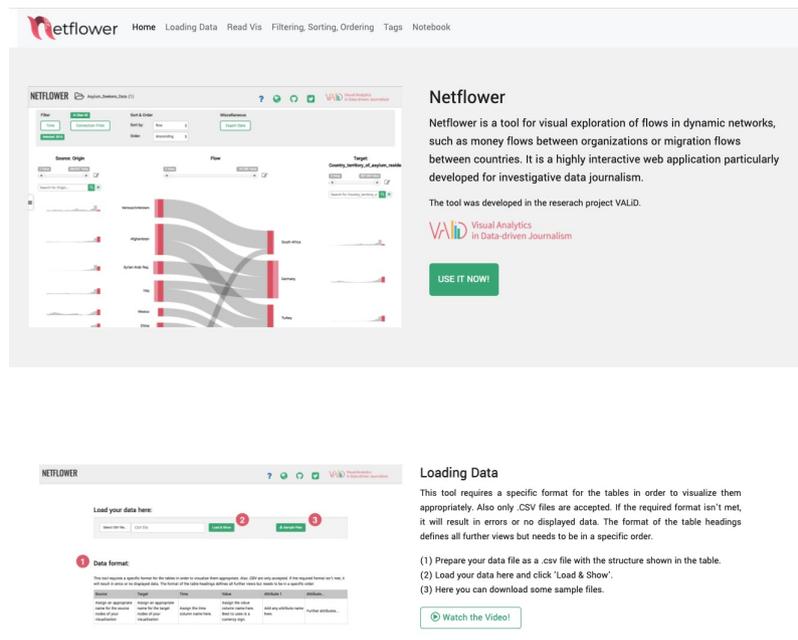

Figure 4: Netflower onboarding consisting of multiple sections including text along annotated screenshots and a video with verbal explanation of the functionality: loading data; reading the visualization; filtering, sorting, ordering; using tags; and notebook.     https://netflower.fhstp.ac.at

**Navigation :** The selected step within  the step-by-step instructions is highlighted in the respective sections (*reading, interacting, using*) as well as in the visualization itself, while the other steps are greyed out (see Figure 2 (left). Users can navigate through the descriptions in a stepwise manner either by using the *Next* and *Previous* buttons on the top right (Figure 2 ⬛) or by direct manipulation through clicking on the numbered textual description within the step-by-step guide. As an alternative for those who do not need a step-by-step explanation, we provide a *Show All* toggle element to hide and show all hints at a glance (see Figure 2 (right).

**Step-by-Step Messages ⬛:** Based on the explanatory sequences of our interview participants, the step-by-step guide incorporates textual descriptions on how to *read, interact with, and use the chart*. The first part contains information, e.g., about the visualizations' shape, axes, or color coding. The second part *interacting* with the chart emphasizes the applied interaction techniques, e.g., indicating how to re-order axes, or filter attribute values. For the part *Using the Chart*, we provide three examples using the low-level typology (identification, comparison, and summarization task) by Brehmer and Munzner [46].

**Visualization with Visual Cues ⬛:** The design is organized in numbered textual descriptions with highlighted attributes in combination with in-place annotations [74] (numbers and symbols) to indicate the connection between the highlighted text elements and the visual encoding. Inspired by different annotation designs by Lu [40], we use circular visual markers with numbers which relate to the selected step. To understand the visual encoding, we highlight *data*, *guides*, and *marks*, as used in the Vega grammar [59]. Hence, in each step, we accentuate words that are related to visual properties of the visualization, an encoding, a data transformation, a pattern, or a finding. Similarly, to explain patterns (e.g., positive or negative correlations between attributes), we use rectangles overlaid in the visualizations as shown in step 3 in Figure 2. In addition, for demonstrating how to interact with the visualization, we introduce small icons, e.g., for brushing and reordering axes . As outlined in Figure 2 (right), we adapted the visual representation of the step-by-step guide and applied it to a scrollytelling tutorial, where users had to scroll from top to bottom to see descriptions and hints within and besides the chart.

*Scrollytelling Tutorial:* Scrollytelling [3] is a powerful narrative format to package and transmit complex information [61, 55]. Based on the principles of construction by Nolan et al. [47], text elements can be displayed incrementally by scrolling up and down the screen [55, pp. 95]. Numbers next to the text elements indicate the current and the total amount of instructions to give users an idea about the length of the scrollytelling. The content and organization (description number and assignment to one of the three sections) is in line with the step-by-step guide to ensure comparability. Our scrollytelling prototype (learning environment cf. [63]) for study 2 shows  the interactive visualization with visual cues (annotations) on the left and ⬛ ten instruction steps on the right side of the user interface, see Figure 2.

As a third method, we introduce video tutorials, and YouTube videos that are commonly used in the context of onboarding and help systems [6, 51, 28, 56, 48, 36, 37].

*Video Tutorial:* Video tutorials rely on passive instructions, as interaction with the visualization is not supported. Similar to the explanation and structure of the step-by-step guide, we show each *mark* and interaction element and describe with a voice-over each step of the explanation. To enable auditory impaired people understanding these explanations, textual descriptions with subtitles are provided.

The step-by-step guide, the scrollytelling tutorial, as well as, the video tutorial can be categorized as [63] "learning environments", independently from of a VA tool. Therefore, we integrated an in-situ (internal onboarding cf. [63]) onboarding into the VA tool *Netflower*, using a scrollytelling approach. Results of Study 2 in Section 5 indicated scrollytelling as an promising approach. In the next paragraph, we introduce the design of the in-situ onboarding in detail.

*In-situ Onboarding using Netflower:* Netflower[1] [64] is an interactive web application for visually exploring dynamic networks. Its landing page already contains an integrated onboarding consisting of annotated screenshots, textual descriptions, and videos. For our comparative evaluation, we compared the already existing onboarding (external onboarding cf. [63]) (shown in Figure 4) with an in-situ developed scrollytelling (internal onboarding cf. [63]) (shown in Figure 3). When scrolling from top to bottom, text elements are highlighted incrementally at the right side of the interface , next to the main visualization , see Figure 3. As the side pane is linked with the interactive visualization, each paragraph is highlighted in red respectively. The scrollytelling behavior in the in-situ onboarding is similar to the scrollytelling tutorial in study 2, as the scrolling guides the user through the instructions. The difference is that the in-situ scrollytelling shows all the instructions in the side pane and triggers the user with highlighting of the respective paragraph. However, the scrollytelling tutorial provides the instructions step-wise without showing all the instructions at a glance.

The text for this in-situ onboarding is the same as in the already existing onboarding integrated in the landing page of the Netflower tool, as shown in Figure 4. The Sankey Diagram is interactive, e.g., by clicking on a connection line in the Sankey diagram, a detail view showing the size of the flow between the two nodes will be displayed. To onboard the user to those features, we simulate the interaction and show the detail view, for example, while scrolling through the onboarding messages.

## 3.2. Implementation

The prototypes used for this contribution are built with web technologies such as JavaScript, HTML, and CSS. We rely on React [52] as the basic front-end framework to provide a responsive user interface. For the visual design, layout, and the explanation steps, we benefit from the styling framework Ant Design [4], as it offers a large variety of user interface elements and guarantees a responsive grid. For rendering the charts, we rely on Vega and the Vega-Lite wrapper for React [53] and D3.js [16] to integrate the in-place annotations. As we faced limitations concerning the supported interaction techniques (e.g., brushing, reorder-ing, and tooltips) with Vega [59] and Vega-Lite [58] while developing the parallel coordinates plot, we modified the library to enable the interactive exploration [33]. For the prototype used in study 3, we relied heavily on the code base of Netflower, which is available under the MIT license at https://github.com/VALIDproject/netflower. We forked it and included our in-situ scrollytelling, without the use of additional libraries. All prototypes can be accessed here: https://onboarding-methods.netlify.app/.

## 3.3. Task Categories

To determine differences regarding task difficulty, we use the model by Friel et al. [23] that distinguishes between three stages of graph comprehension: 1) *reading the data*, 2) *reading between the data* and 3) *reading beyond the data*. As outlined by Lee et al. [39], *reading beyond the data* can only be assessed using open-ended questions with qualitative answers. We designed tasks for all three studies following the Visualization Literacy Assessment Test (VLAT) [39], which defines a set of questions for tasks such as retrieve value, find extremum, determine range, or make comparison for 12 different visualization types. However, not all visualization types used within this contribution are covered by the VLAT. Therefore, we adapted questions for not yet described visualization types (e.g., parallel coordinates plot) by using the VLAT questions as a template. All task questions can be retrieved from the url https://phaidra.fhstp.ac.at/o:4841. Examples for the parallel coordinates plot are the following:

*Retrieve Value*: What is the percentage of cacao in Palette de Bine?
*Find Extremum*: Which company had the highest sales?
*Determine Range*: What was the average price per kg for cacao beans with a 70% cacao percentage?
*Make Comparison*: Comparing Caribeans, Kah Kow, and Sibu, which of these companies shows the lowest price per kg and the highest sales volume (in Euro)?
*Correlation*: In general, there is a (negative | positive) correlation between the review date and the sales volume.
*Similarity*: What are the most similar companies in 2015 regarding the rating and price per kg?

Further and in line with [38, 35, 22], we use easy-to-understand, concrete, and time-oriented data sets, see Sections 4.3 and 6.4 for a detailed description of the used data sets in each study.

## 4. Study 1: Visualization Onboarding and Effect on User Performance

As outlined in Figure 1, we designed study 1 to understand how visualization onboarding affects the users' performance (**RQ1**) using different visualization types (**RQ2**) and task difficulty levels (**RQ-A**). We therefore analyzed four visualization types (bar chart, horizon graph, change matrix, and parallel coordinates plot), with two conditions (with and without a step-by-step guide as onboarding method) and with interchanged data set for pre- and post-tasks, resulting in 16 conditions. Consequently, we report on differences in the performance measures answer correctness and improvement (in percent), response time and response time improvement (in seconds), the interaction time with the onboarding part (in seconds) and interaction patterns as well as qualitative feedback on the onboarding approach. Thus, we followed a mixed-method approach to increase the comprehension of visualization onboarding usage.

---

[1] https://netflower.fhstp.ac.at

### 4.1. Participants and Apparatus

We conducted 16 experiments on the crowdsourcing platform Amazon Mechanical Turk. Each worker was compensated with $3 for surveys with onboarding and $1.5 for surveys without onboarding. Furthermore, we required the workers to have at least a Bachelor's Degree and fulfill the MTurk definition of a Master, meaning that they consistently showed good survey results. Due to the fact that we released the surveys sequentially, we excluded workers if they had already participated earlier. In total, we recruited 400 MTurk workers (25 workers per condition), wherefore we had to exclude 12 due to an incomplete survey. Thus, 388 MTurk workers (Gender: m = 241, f = 149, prefer not to say=2); Age: $M = 38.23$, $SD = 9.60$) participated in our survey.

We conducted the experiment using LimeSurvey[2] where we made use of the time recording and question randomization features. We designed and implemented the visualization for the pre- and post-tasks and onboarding parts as described in Section 3.2. We linked the onboarding part to LimeSurvey and integrated the visualization parts (pre- and post-tasks) into the questionnaire using iFrames. To assess interaction time with the onboarding concept based on videos and click events, we used the behavior analytics tool Hotjar[3]. Participants used their own devices. Since our onboarding concept was designed for desktop and mouse-based devices only, we excluded the usage of mobile devices not only in the textual description but also programmatically. Two of the co-authors inspected the videos independently for identifying different aspects to answer the following questions: "How long do users on average interact with the onboarding concept? Was there any interaction with the visualizations? Which interaction behaviors can be observed?"

### 4.2. Procedure

Figure 5 shows both survey conditions, one with and the other without visualization onboarding. Both surveys were structured equally, except the one with onboarding incorporates the interactive onboarding part and seven questions about the ease of use, confidence, comprehensibility, style, interpretation, interaction, and frequency of use of the onboarding part itself. Each survey started with a question about the level of experience for the given visualization type by using a 5-point Likert scale. In line with previous work [10], the bar chart shows the highest average level of experience (71.7%), followed by the horizon graph (34.0%), the change matrix (25.3%), and the parallel coordinates plot (22.2%) (see Figure 6). After the assessment of the experience level, we continued with the pre-task set: four questions each for bar chart, horizon graph, change matrix, and six questions for parallel coordinates plot. The questions were posed in randomized order to avoid a selection bias and as outlined in Section 3.3, the questions were inspired by the VLAT. Next, participants were obliged to interact with the onboarding part—if assigned to the onboarding group. To ensure that they actually interacted with the onboarding, we added an obligatory question of confirmation before continuing with the survey. This was followed by the set of questions of the post-tasks. Finally, we completed the survey with questions about demographics (e.g., age, gender, profession).

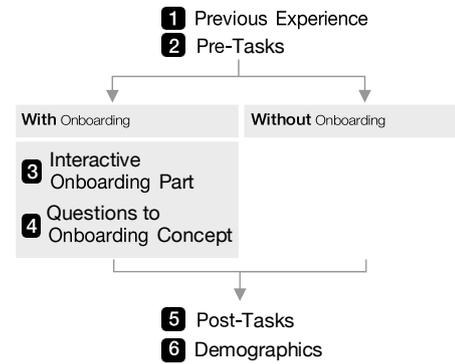

Figure 5: Procedure of study 1 and study 2 with MTurk workers. The surveys followed a uniform structure with two conditions—"with" and "without" onboarding—where the "with" onboarding condition included an interactive onboarding part in addition to a questionnaire to gather feedback about the onboarding concept.

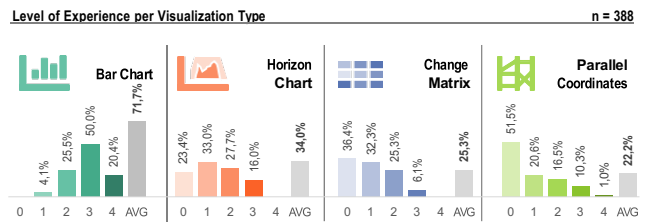

Figure 6: Level of experience per visualization type for study 1 on a 5-point Likert scale, where 0 means no experience at all and 4 highly experienced. The percentage value indicates how many participants subjectively rated their respective level of experience. Additionally, the 'AVG' bar outlines the average level of experience per visualization type from 0-100%.

### 4.3. Data Sets

To follow the concept of easy-to-understand and concrete data sets, we used weather, car data, and Olympic medals distributions for study 1. For the weather data we have chosen the publicly available data set about the daily 20th-century surface air from the European Climate Assessment & Data set project [17]. We calculated the average temperature of European cities for selected years (1990, 1991, 2018) using this data set for the bar chart and horizon graph onboarding. For the change matrix, we used the Olympic data set [49] showing the medal distributions between 1990 and 1991. The data sets for the parallel coordinates plot required more dimensions. Thus, we decided to take the car data set from PACO [13] for our visualization onboarding which we extended by the production year of the cars in order to add the time aspect. For the pre- and post tasks we used easy to understand, time-oriented data sets (e.g., Spotify data based on song titles and various characteristics [62], chocolate bar rankings [15], amount of steps over time, weather data [17], and Olympic medals [49] distribution over time).

| Performance Measure | Answer Correctness (in %) | | Response time (in s) | |
|---|---|---|---|---|
| Onboarding | With | Without | With | Without |
| **Bar Chart** | $0.88_{.33}$ | | $30.25\ s_{.26.44}$ | |
| | $0.87_{.33}$ | $0.88_{.17}$ | $33.92\ s_{.28.02}$ | $26.58\ s_{.24.31}$ |
| **Horizon** | $0.89_{.32}$ | | $38.26\ s_{.28.21}$ | |
| | $0.90_{.302}$ | $0.87_{.34}$ | $38.86\ s_{.27.80}$ | $37.65\ s_{.28.69}$ |
| **Change Matrix** | $0.76_{.43}$ | | $43.73\ s_{.72.61}$ | |
| | $0.81_{.39}$ | $0.70_{.46}$ | $52.78\ s_{.95.56}$ | $34.50\ s_{.34.49}$ |
| **Parallel Coordinates Plot** | $0.55_{.50}$ | | $57.21\ s_{.77.00}$ | |
| | $0.52_{.50}$ | $0.58_{.49}$ | $65.91\ s_{.61.70}$ | $50.13_{.99.60}$ |

Table 1: Performance measures for all four visualization types of study 1, n=388.

### 4.4. Results

*Visualization Type:* We started by assessing the differences between all four visualization types. We identified significant differences in the answer correctness between the four visualization types. The highest answer correctness for post questions was given for the horizon graph ($M = .89; SD = .32$), followed by the bar chart ($M = .88; SD = .33$), the change matrix ($M = .76; SD = .43$), and the parallel coordinates plot ($M = .55; SD = .50$), $F(3, 1742) = 70.557, p = .000$. With regard to the response time, the fastest answers were given for the bar chart ($M = 30.25\ s; SD = 26.44$), then the horizon graph ($M = 38.26\ s; SD = 28.21$), the change matrix ($M = 43.73\ s; SD = 72.61$), and then the parallel coordinates plot ($M = 57.21\ s; SD = 77.00$) survey, $F(3, 1743) = 18.523, p = .000$. The results based on an answer correctness or response time improvement reveal that statistically significant differences between the visualization types exist for the answer correctness improvement, $x^2 = 6.643, p = .084, df = 3$. In contrast, for the response time improvement, participants answered post questions faster with the horizon graph ($M = -20.58\ s; SD = 234.42$), then with the parallel coordinates plot ($M = -19.32\ s; SD = 95.01$), followed by the bar chart ($M = -2.315\ s; SD = 42.15$) and the change matrix ($M = 1.63\ s; SD = 100.09$) — irrespective of the onboarding concept, $F(3, 1742) = 3.190, p = .23, tl^2 = .005$. The more negative the value is, the faster participants answered the post question set. Thus, it can be said that questions about horizon graphs and bar charts could be answered more correctly than the other visualization types but the average response time, for instance for the horizon graph, was longer for the post- than for the pre-question set.

*Overall Onboarding:* We continued by analyzing the differences of the measures between a step-by-step onboarding guide and our control group without any onboarding at all. We hypothesized that the answer correctness is higher for surveys with onboarding than without any onboarding. Using all surveys from all visualization types, we could not determine statistically significant differences between the answer correctness between surveys with ($M = 879.25$) or without onboarding ($M = 867.85$), $U = 376064.00, z = -.624, p = .533$. In contrast, the participants conducting surveys in the control condition showed statistically significant faster answers ($M = 38.74\ s; SD = 63.43$) than participants in the onboarding condition ($M = 49.73\ s; SD = 56.41$), $t(1742) = 3.821, p = .000, eta^2 = 0.008$. The improvement measures did not show significant differences between both conditions. Thus, across all visualization types, we could not identify differences or improvements.

*Onboarding per Visualization Type:* We therefore decided to investigate if the utilization of an onboarding has an influence on the measures for each visualization type independently. We used a one-way multivariate analysis of variance (MANOVA) with answer correctness and response time as dependent variables and the onboarding concept as well as the visualization type as fixed factors. There was a statistically significant difference in user performance based on the onboarding concept for each visualization type, $F(6, 3470) = 2.607, p = .016; Wilk's A = .991, tl^2_p = .004$. However, only for the answer correctness ( $F(3, 1736) = 3.310; p = .019; eta^2_p = .006$ ) and not the response time ( $F(3, 1736) = 1.798; p = .145; eta^2_p = .003$ ). Regarding the performance improvement, no significant differences could be observed, neither in the answer correctness improvement, nor for the response time improvement. To investigate the effect further on visualization type level, a significant difference in the performance could be observed for the bar chart, $F(2, 389) = 3.835, p = .022; Wilk's A = .981$. However, only for the response time ($F(1, 392) = .7.662; p = .006$) between the two conditions but not for the answer correctness ($F(1, 392) = .094; p = .759$). This means that surveys without any onboarding were answered faster $26.58\ s \pm 24.31$ than surveys with an onboarding concept ($33.92\ s \pm 28.02$). For the horizon graph, neither in the answer correctness, nor in the response time, any performance differences could be determined between surveys with and without onboarding, $F(2, 373) = .485, p = .616; Wilk's A = .997$. In contrast, for the change matrix, both performance measures show statistically significant differences, $F(2, 393) = 6.698, p = .001; Wilk's A = .967$. Interestingly, the answer correctness was on average higher for surveys with an onboarding ($0.81 \pm .39$), where the condition without any onboarding shows a lower answer correctness ($0.70 \pm .46$), $F(1, 396) = 6.675; p = .010$. Additionally, with regard to the response time, surveys with onboarding ($52.78\ s \pm 95.56$) were answered on average slower than surveys without onboarding ($34.50\ s \pm 34.49$), $F(1, 396) = 6.357; p = .012$. In summary, participants using the change matrix invested more time with the visualization and achieved a higher answer correctness. Lastly, also for the parallel coordinates plot, a statistical difference in the performance between surveys for the two conditions could be assessed, $F(2, 577) = .3.506, p = .031$; $W ilk^t sA = .988$. However, only the response time shows differences, $F(1, 580) = 6.048; p = .014$, where surveys without onboarding ($50.13 s \pm 99.57$) showed on average faster response times than surveys with onboarding ($65.91 s \pm 41.70$). The answer correctness without onboarding showed surprisingly more correct answers ($0.58 \pm .49$) than for surveys with the onboarding ($0.52 \pm .50$), but with a large variance.

*Onboarding and Task Difficulty* Subsequently, we investigated the influence of the question type difficulty the users' performance outcome. We therefore assessed differences between *reading the data*, *reading between the data*, and *reading beyond the data*. Irrespective of the onboarding method and the visualization type, *reading the data* questions showed the highest mean rank (*meanrank* = 955.13), followed by *reading between the data* (*meanrank* = 813.33) and *reading beyond the data* question types (*meanrank* = 803.58), $x^2 = 64.044, p = .000, df = 2$. Similarly, the response time is in accordance with the answer correctness, where the fastest answers were given for the *reading the data* ($M = 28.0305 s; SD = 34.70184$), then for the *reading between the data* ($M = 55.25 s; SD = 48.46$), and lastly for the *reading beyond the data* questions ($M = 58.82 s; SD = 88.22$), $F(2, 1741) = 53.651, p = .000, tI^2 = .058$. An additional Tukey HSD post-hoc test reveals that a statistically significant difference between *reading the data* and both *reading between* and *reading beyond the data* question type ($p = .000$) but not between *reading between the data* and *reading beyond the data* ($p = .610$) over all visualization types exists.

Analyzing the performance measures using a three-way ANOVA between onboarding concept, visualization type, and task difficulty level shows statistically significant differences in the response time but not for the answer correctness, $F(6, 1722) = 1.485, p = .180, tI^2 = .005$. Interestingly, with regard to the response time improvement, differences could be determined, $F(6, 1722) = 6.016, p = .000, tI^2 = .021$. Based on the response time improvements, we would argue for a learning effect for all visualization types for the participants that did not receive an onboarding, as all question types (Friel et al.) were answered faster ($M = -15.07 s; SD = 188.30$) for the post questions rather than the pre question set. In contrast, participants who were in the onboarding condition show diverging results. In particular, the *reading between the data* question ($M = 57.12 s; SD = 118.31$) for change matrices with onboarding took statistically significant longer to answer than all the other question types ($M = 22.33 s; SD = 488.90$). Moreover, the *reading beyond the data* questions ($M = -55.06 s; SD = 137.53$) for the parallel coordinate plots were answered faster by participants that received an onboarding ($M = -23.53 s; SD = 105.65$). This goes along with the feedback as "Once I read the step-by-step guide, using the graph felt way easier" (St55).

Similarly, the *reading between the data* task for horizon graphs was answered faster from participants that had an onboarding to the visualization type.

## 4.5. User Behavior

We used the recordings of mouse and click events of the participants' sessions to investigate the interaction with our onboarding concepts. Due to blocking software in browsers or other available options which protect users from tracking or recording their actions, it was only possible to collect interaction data from 91 out of 145 participants across all studies who were supposed to do the onboarding.

*Interaction Time* Based on the recorded videos, we analyzed the *interaction time* by checking when the user stopped interacting with the onboarding prototype. Interaction time is defined as *no mouse movement any longer*. We found that the MTurk workers interacted with the onboarding concept independently of the visualization type ($M = 102.4 s; SD = 213.71$). Interaction time with bar chart ($M = 65.96 s; SD = 75.40$) and horizon graph ($M = 60.19 s; SD = 66.59$) was moderately low. In contrast, interaction time for change matrix ($M = 204.05 s; SD = 446.41$) and parallel coordinates plot ($M = 99.86 s; SD = 96.33$) was higher.

*Interaction behaviors* Extracting information from the mouse tracking data helped us to determine hot spots on the onboarding concept. We evaluated these hot spots as attentive areas with our four areas of interest: **A** introduction, **B** navigation, **c** step-by-step guide, and **D** visualization (Figure 2). As the area for *introduction* does not contain interactive elements, hardly any mouse movement and no click events could be observed. When analyzing patterns on the visualization itself, we could observe that the hover feature was frequently used for all visualization types. In particular when the step-by-step guide referred to attributes or exact measures from the data set. Interestingly, we found that besides the mouse movements, more click events were ascertained with the parallel coordinates plot. This can be traced back to the introduced interaction possibilities, e.g., brushing and re-ordering axes (see Figure 7). Sessions from the bar chart, horizon graph, and the change matrix show that the majority of participants interacted with both the navigation pane and the step-by-step guide in a balanced manner. In contrast, for the parallel coordinates plot, we were able to distinctly observe that users predominantly clicked on the text within the step-by-step guide. Hardly any superimposed mouse movements were detected with our navigation pane. Moreover, when assessing the usage of the *Show All* button, we recognized that it was only used by 10 out of 91 participants (5.08%) for all visualization types. Caused by the active usage of the step-by-step instructional descriptions, we analyzed the explanation sections in further detail. We noticed less interaction or interest in the *Using the Chart* section than with *Reading the Chart* and *Interacting with the Chart* indicated by fewer mouse movements or click events, and by a lower retention time. This applies to all visualization types. Particularly noteworthy was this behavior for the parallel coordinates plot which had the additional section for interaction and thus more steps. We noticed that 42 participants out of 56 total records for parallel coordinates plot went much faster over the *Using the Chart* section or completely ignored it. It could be noted across all studies that this section seemed to be less interesting to the participants.

Figure 7: Mouse movement heatmap on the step-by-step guide for parallel coordinates plot received from Hotjar.

Overall, it can be said that the familiarity with a visualization type has an impact on the answer correctness. Thus, for the rather unfamiliar visualization types, the change matrix and the parallel coordinates plot, it can be seen that onboarding is needed to improve the performance. In contrast, for highly familiar visualization types, an onboarding is not required and has no effect on the performance measures. We therefore continued to investigate different types of onboarding concepts in study 2 using the parallel coordinates plot.

## 5. Study 2: Types of Visualization Onboarding Methods and the Effect on User Performance

As we identified a need for onboarding for the more unknown visualization type—the parallel coordinates plot—we aimed to observe differences between the means how onboarding can be provided to a user. We therefore implemented and assessed three methods: (i) an interactive step-by-step guide, (ii) a scrollytelling tutorial, and (iii) a video tutorial with voice-over. Tutorial and videos are common ways to support users. Hence, we wanted to explore scrollytelling as a new approach to the more traditional concepts of a tutorial and videos. We build upon the previous study and compare the measures for the parallel coordinates plot. Thus, we aim to understand how the different onboarding methods affect the users' performance (**RQ3**) and if a difference of user performance can be assessed for different task types (**RQ-A**). Further, we applied a sentiment analysis on the qualitative feedback to understand differences in the perception of the three onboarding methods. We therefore made use of the sentiment analyser by MonkeyLearn [42] — a machine learning platform that aims to retrieve and classify text information — while also evaluating each qualitative feedback ourselves. We used the same study design as for study 1. The results of the sentiment analysis are already published by Stoiber et al. [65].

### 5.1. Participants and Apparatus

We conducted a between-subject design and employed additionally 100 MTurk workers, where 25 were assigned to each condition for the yet unknown onboarding methods (scrollytelling and video; changed data set positions). In total, we had to exclude 5 participants, hence answers from 145 participants (m = 94, f = 50, prefer not to say=1; Age: $M = 35.46$, $SD = 9.29$) were additionally taken into account. As Hotjar builds heatmaps on top of screenshots, we were not able to assess the scrolling behavior and thus did not assess mouse movements for the method comparison.

### 5.2. Results

*Overall Onboarding:* First, we started to determine differences in the performance measures (answer correctness and response time) for all four conditions (the baseline without onboarding, step-by-step guide, scrollytelling tutorial, and video guide). No statistically significant difference in the answer correctness for the post question set could be determined. However, the response time showed that faster answers were given after a scrollytelling tutorial ($M = 41.17 s$; $SD = 46.55$), followed by the step-by-step guide ($M = 49.97 s$; $SD = 59.77$), the condition without any onboarding ($M = 58.07 s$; $SD = 104.90$), and the video tutorial ($M = 58.39 s$; $SD = 71.47$), $F(3, 1134) = 3.3785$; $p = .010$; $Eta = .096$.

*Onboarding and Task Difficulty:* For the parallel coordinates plots, differences in the task complexity level could be observed for both performance measures (answer correctness: $F(2, 1134) = 37.081$; $p = .000$; response time: $F(2, 1134) = 36.66$; $p = .010$). A Kruskal-Wallis test showed that the highest mean rank for the dichotomous variable answer correctness was achieved by the easiest task (669.00), followed by the intermediate (550.50) and advanced task (501.00), $H(2) = 69.33$, $p = .000$. As expected, an ANOVA with post-hoc SNK test showed that *reading the data* tasks were answered the fastest ($M = 30.918 s$; $SD = 64.53$), followed by *reading between the data* $M = 49.76 s$; $SD = 56.01$) and *reading beyond the data* $M = 74.79 s$; $SD =$

| Task Difficulty* | Answer Correctness (%) | | | Response Time | | | Sentiment Analysis** | | | | | | | |
|---|---|---|---|---|---|---|---|---|---|---|---|---|---|---|
| | 1 | 2 | 3 | 1 | 2 | 3 | Positive | | Negative | | Neutral | | No answer | |
| **Baseline** No Onboarding | .71±.46 | .52±.50 .46±.50 | .39±.49 | 33.56 s±30.53 | 58.07 s±104.90 54.19 s±58.87 | 86.45 s±165.59 | - | - | - | - | - | - | - | - |
| **Step-by-Step Guide** | .74±.44 | .52±.50 .47±.50 | .35±.49 | 27.11 s±34.26 | 49.97 s±59.78 48.20 s±52.78 | 74.61 s±75.45 | 13 | 27.66% | 3 | 6.38% | 4 | 8.51% | 27 | 57.45% |
| **Scrollytelling Tutorial** | .65±.48 | .42±.50 .47±.50 .42±.50 | | 28.57 s±28.76 | 41.17 s±46.55 | 53.85 s±54.99 | 16 | 32.65% | 6 | 12.24% | 9 | 18.37% | 18 | 36.73% |
| **Video Tutorial** | .60±.49 | .49±.50 | .37±.49 | 34.44 s±42.62 | 58.39 s±71.47 55.75 s±62.45 | 84.98 s±91.73 | 21 | 42.86% | 1 | 2.04% | 6 | 12.24% | 18 | 42.86% |

\* 1= Reading the data; 2= Reading between the data; 3 = Reading beyond the data

\*\* Total indicates the number of submitted positive/negative/neutral feedback in relation to the total number of participants assigned to a method.

Table 2: Overview of performance measures for all four onboarding conditions (baseline with no onboarding, step-by-step guide, scrollytelling, and video tutorial) and three task difficulty levels. Additionally, the results on the text classification using the sentiment analysis for study 2 are presented. *Note*: Not all participants answered these questions as they were not mandatory.

104.99). However, similar results for each onboarding technique could be observed, see Table 2.

*Qualitative Feedback:* Overall, the results show that 34.48% of the responses can be classified as positive, 6.90% as negative, 13.10% as neutral statements, and 45.52% of the participants did not submit any feedback on the onboarding as it was not mandatory. Participants decisively appreciated the condensed, structured, and grouped explanation steps of each of the approaches. On closer examination, the highest positive feedback was given for the video tutorial (42.86%), followed by the scrollytelling tutorial (32.65%), and the step-by-step guide (27.66%). Noteworthy is that participants highlighted learning new features during the video tutorial (V26, V42, V50), e.g., *"The video was helpful and showed me some features that I wasn't familiar with [..]"* (V50). More specifically, *"[..] I liked knowing that I could move columns next to one another, etc."* (V60), which relates to the re-arrangement of axes. However, the automatically generated voice-over in the video guide was described as unattractive as it sounded robotic (V3). In contrast, the step-by-step guide did not support the usage and understanding of interactive elements (e.g., filtering and moving axes) or the interpretation of correlations as *St34* described: *"I have trouble with correlations, but I don't think that is the fault of the guide—although examples would be good"* (St34). This may also explain why the results of study 1 were not significantly improved for the onboarding condition as anticipated. Regarding the scrollytelling tutorial, users *"[..] enjoy that the images on the left do not appear until I need to see them, which prevents confusion"* (Sc38). Hence, the display of information on-demand through scrolling and thereby enriching the visualization with information increased the perceived satisfaction of the participants.

Although statistically no significant differences can be found in the different onboarding methods, the qualitative feedback from the participants supports the statement of needing an onboarding predominantly for unfamiliar visualizations. In addition, the video tutorial shows a high resonance with 42.86% positive feedback, since the approach of storytelling, but also of guidance can be adopted here. As the results in Table 2 shows that the average answer correctness for the scrollytelling tutorial was rather low (but with high variance), we decided to investigate the effect of onboarding even more with study 3, where onboarding was embedded in a software. Furthermore, we wanted to investigate new ways of support users by using the scrollytelling approach.

## 6. Study 3: In-situ Visualization Onboarding in Netflower

Our aim in this last study was to compare an external and an in-situ onboarding method (see Figure 1). We added in-situ onboarding to an existing VA tool, to see if differences in performance (**RQ4**) and attitude (**RQ5**) of users exist compared to an external onboarding. We use the term *attitude* as it is used in psychology to describe a user's way of thinking about something, and their corresponding behaviors, beliefs, and emotions towards it [25]. Inspired by the results of the previous studies (study 1 & 2), we additionally investigated if we can detect any difference with regards to tasks of different complexity as described by Friel et al. (**RQ-A**). In the following, we present the study design, participants, apparatus and material, as well as describe the procedure and the results of our comparison.

### 6.1. Study Design

The study was performed in two rounds, the first taking place in spring 2020 and the second in fall 2020, illustrated in Figure 8. In each round the students were split into two groups. Each group completed the experiment in two sessions, with three weeks in between each session, and using a different data set each session. In the first round one group was assigned the external onboarding, and one group was assigned the in-situ onboarding. They used the same onboarding for both sessions, to establish the existence or absence of a learning effect. In the second round, one group used the in-situ onboarding for the first session, and the external onboarding for the second session, whereas the second group did the opposite, to enable within-subject comparison. The experiment was always the same in each session: The participants were given a survey, which included an introduction as well as different tasks, and a questionnaire to be answered after completing those tasks.

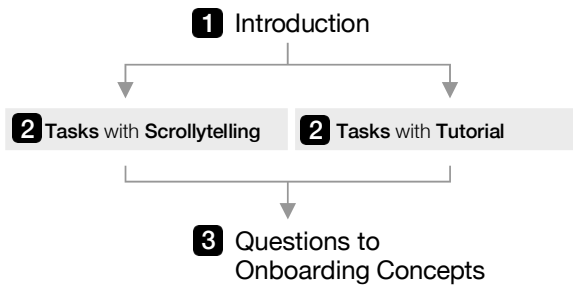

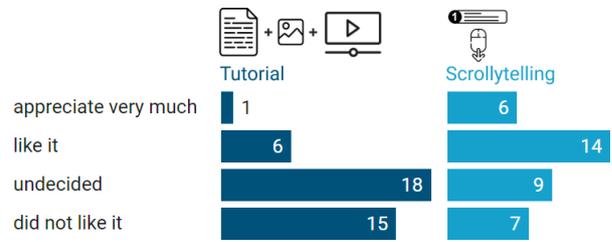

Figure 9: Participants' attitude towards onboarding concepts. *Note:* Not all participants answered these questions as they were not mandatory.

Figure 8: Procedure for study 3 with students. The flowchart shows the structure of the experiment. The table shows how we split the participants into groups A-D, which groups were assigned which onboarding system, and how many people participated in parentheses.

|  | Spring 2020 | | Fall 2020 | |
|--|-------------|---|-----------|---|
|  | Scrollytelling | Tutorial | Scrollytelling | Tutorial |
| Session 1 | A (13) | B (13) | C (18) | D (21) |
| Session 2 | A (13) | B (11) | D (20) | C (18) |

## 6.2. Participants and Apparatus

We recruited 65 students (Gender: m = 42, f = 23, prefer not to say=0) from the first year of the international Master's programs "Data Science" and "Media Informatics" at the TU Wien, where the study was conducted in the context of a lecture. The participants were required to have at least basic knowledge about data visualizations. For round one, 24 participants completed both sessions, with an additional two that only completed the first one. In round two, 39 participants completed both sessions, but 2 participants obviously collaborated in the second session, so the results of only one was taken for further analysis. The exact distribution of participants in each session and round is shown in Figure 8.

In line with the other two studies, we used LimeSurvey to present the tasks and the questionnaire. The participants had to open the prototype (described in Section 3) in parallel to the survey.

## 6.3. Procedure

The survey started with an introduction about the study and the Netflower tool. The participants were given a scenario in which they should imagine themselves as data journalists and asked to complete 14 tasks. The first 11 tasks were presented as multiple-choice aimed for a distinct answer, e.g. "Compare Malaysia, Sweden and Austria in 2016. Which country received the most asylum seekers?" In contrast, the last three questions were posed as free-text answers to understand the analysis behavior, e.g. "Compare the development of asylum seekers between 2000 and 2016 between Austria,

Germany and Sweden. Is the development in Austria more similar to Sweden or to Germany? Explain why you think that this is the case?" The tasks were chosen based on VLAT [39], as discussed in section 3.

After the tasks, they were given a questionnaire of six to nine questions, part multiple-choice and part open-ended, that aimed to make the participants reflect and give feedback. Some questions referred to the usage of the onboarding system, some referred to the quality of the onboarding system and others were more open ended. The final questions asked for the user experience of the participants concerning the onboarding system. The whole survey can be found here: https://phaidra.fhstp.ac.at/o:4872.

## 6.4. Data Sets

Netflower is a visual exploration tool for investigating flows between networks over time, see Figure 3. For our study, we used data about asylum seekers published by the UNHCR (the UN refugee agency) that shows asylum applications from the years 2000 to 2016 in 44 different countries [68] as well as media transparency data about money flows between government entities and media institutions. The media transparency data set is compiled by the Rundfunk und Telekom Regulierungs-GmbH and records money given to media institutions by Austrian government entities [45]. In each first session, the asylum seekers data set was used, in each second session the media transparency data set was used.

## 6.5. Results from the Questionnaire

The questionnaire consisted of two parts, one part containing multiple choice questions and another part consisting of open-ended questions.

*Multiple-choice questions:* When asked whether they started to study the onboarding system, or on the other hand, they started to explore the system on their own, almost 50% of the participants reported that they read the text in the scrollytelling system first while only about one fifth of the participants in the tutorial group did. We could not detect any difference with regards to the intensity of usage of the onboarding system: In both conditions about one fifth of the participants reported that they used the onboarding system "often" and about 50% of them reported to have "never" used it at all. The rest reported to have used the system "some-times". We were also interested in the acceptability of the amount of information displayed, therefore, we asked if there was too much text in the onboarding system. Again, no differences between the two conditions could be determined. For both methods, about one third claimed that there was too much text, the rest did not agree with this statement. For the scrollytelling condition, we additionally asked about the helpfulness of the gradual appearance of the elements, which was perceived by about 73% as a helpful. Additionally, the simple introduction into the system was mentioned to be supportive.

*Open-ended questions:* The open ended questions were grouped together and are treated as one question that can be assigned to multiple categories. The questions that had more varied responses were further evaluated using qualitative content analysis [44, 60] and discussed here. The aim of a qualitative content analysis was to analyse textual or other material in a systematic way. It is applied when the analysis process does not yield entirely obvious categories and extended interpretation of the material is necessary. The analysis process consisted of repeatedly coding the material and results in a few categories that describe the most important insights that can be derived from the material.

In this part of the questionnaire, the question whether participants appreciated the onboarding system (either in-situ scrollytelling or tutorial) yielded interesting results. The participants were specifically asked to discuss advantages and disadvantages. Therefore, the answers tend to include both positive and negative aspects.

In general, participants liked scrollytelling more than the tutorial. Six participants appreciated scrollytelling very much, 14 liked it, nine were undecided, and seven did not like it at all. In contrast to that, only one participant liked the tutorial very much, six appreciated it, 18 were undecided, and 15 did not like it. Note that the open-ended questions were not answered by all participants as they were not mandatory.

The following statements describe the positive attitudes concerning scrollytelling. "I liked it because it visually highlighted the important parts and I was less overloaded with all the possibilities in the beginning, because it showed me one feature at a time" (P3). "I like that it is quick and that it doesn't require too much time to be learned and understood. I liked that the text was not too invasive, being on the side and having the possibility to hide it" (P16).

The most important positive aspects of the system that were mentioned in the answers were: (1) It visually highlights the most important parts and shows one feature at a time (11 mentions); in this way, it also provides an inherent structure of the system. (2) It is directly inside the system, therefore, using it does not disrupt the workflow (7 mentions). (3) It is a simple introduction into the system (13 mentions).

Participants also outlined drawbacks of the system: (1) The most important negative aspect was that there were technical and usability problems (10 mentions). The scrollytelling system was a prototype, and sometimes usability issues occurred. A few of the participants mentioned that they did not need scrollytelling and discovered the features of the sys-tem themselves (5 mentions). One example for a negative attitude concerning scrollytelling is the following: "I did not really like it. The system to use is quite intuitive, and the scrollytelling is in comparison too long and convoluted. At the same time, it did not cover enhanced topics like notes and tags. I prefer an approach where one starts immediately using it, and later gets the possibility to learn more details" (P12).

The tutorial was appreciated less than the scrollytelling method. Nevertheless, it was stated that "I liked the tutorial because it allowed me to start working quickly. Whenever I struggled, I could go back and read a little more" (P22). The following positive features of the tutorial were mentioned: (1) The tutorial is easy to comprehend, it is nice as a reference and a good overview (15 mentions). (2) Some of the participants also mentioned that they liked the videos (8 mentions).

Several participants had negative attitudes concerning the tutorial: "The tutorial was not really needed because the visualization was quite intuitive and one could guess what functions and filters were available by trying them out. Still, the tutorial gives a good overview about what tools are available" (P11). In the following, we present the most important negative remarks: (1) Many of the participants mentioned that the tutorial was not needed and that they just started working (8 mentions). This is especially interesting because this was not the case with scrollytelling. The explanatory text was the same, but participants invested much more time in interacting with scrollytelling than with the tutorial. Apparently, scrollytelling is more motivating than the tutorial. (2) Several of the participants also mentioned that the tutorial only explained simple things and did not address the more complicated features of the system. They would also have wanted sophisticated examples (9 mentions). (3) Very few participants also mentioned that the did not like videos (3 mentions). Apparently, the participants' attitude towards videos is ambiguous. Some like them very much, and others do not like them at all.

## 6.6. Results from Tasks

We assessed answer correctness (in percent) on multiple choice questions, as well as open ended questions. As study 1 found significant differences in answer correctness based on task difficulty as categorized by Friel et al. [23], we also checked if this holds true for study 3. For the analysis of the free-text questions we defined two main categories:

*Category 1*: Answers, where the correct usage of the system is apparent, for example by having the correct answers or by having written descriptions of how the system was used.
*Category 2*: Any other answers, for example, wrong answers, no answers, answers not rooted in the visualization. An example of an answer for task 12 that was judged to be this category would be "Austria is more similar to Germany due to the geographic location". This could be a valid argument, but by itself does not show that they used the system correctly to come to this conclusion.

We assigned these categories to the free-text answers and evaluated them like the multiple-choice answers (in percent).

*Overall Comparison* Comparing all results of participants using the scrolltelling approach for the first time ($n = 51$, $M = .72$; $SD = .159$) with those using the tutorial for the first time ($n = 52$, $M = .75$; $SD = .159$), we can see no difference in answer correctness ($p = .317$). Here, we use data from session one in round one, as well as both sessions in round two.

*Learning Effect* The goal in the first round (in spring 2020) was to establish the properties of our test setup. We wanted to know how the data set influences responses, as well as how well the participants retain their skills of using Netflower after a month. Comparing the answers of session 1 ($n = 26$, $M = .77$; $SD = .103$) and session 2 ($n = 24$, $M = .78$; $SD = .116$) in spring 2020, we can see that there is no appreciable difference in answer correctness ($p = .878$).

*Within-Subject Comparison* Looking at only the answers in round two, we can compare within subjects. However the answers of participants using the in-situ onboarding ($n = 38$, $M = .71$; $SD = .171$) compared to those using the external one ($n = 39$, $M = .74$; $SD = .176$) show no appreciable difference in answer correctness either ($p = .517$).

*Task Difficulty* We could not find a significant difference in answer correctness between the external onboarding and the in-situ scrolltelling, when grouping by task difficulty. A $t$-test reveals $p = .343$ for elementary tasks, $p = .483$ for intermediate tasks and $p = .598$ for comprehensive tasks. When analyzing the task difficulty using the taxonomy by Lee et al. [39], we noticed that the taxonomy only comprises fairly simple interaction processes. The added additional tasks based on the model by Friel et al. [23] include more complex interaction processes as, for example, relating values from different visualizations in Netflower and then deriving conclusions from this process. These task posed some problems to inexperienced users despite the fact that information about these interaction processes could be found in the onboarding systems. In accordance with the model developed by Friel et al. [23], the simplest cognitive processes (identifying simple values) were easiest. Comparison processes were more difficult, and the most difficult cognitive processes were "compute derived value" which forced participants to "go beyond the data". In general, we noticed that participants made a considerable number of mistakes as in some cases, more than 60% of the answers were incorrect. This contradicts the subjective impression of some of the participants that they were able to understand Netflower without studying the onboarding system and their confidence in the correctness of the results.

# 7. Discussion and Lessons Learned

Introducing users to a new visual layout and guiding them to a higher visualization literacy is the fundamental goal of visualization onboarding. However, as data sets are increasing in complexity, well-known conventional business charts (e.g., line charts, bar charts, and pie charts) are no longer sufficient and new forms of visualization—especially designed to deal with this increasing complexity—need to be applied. The lack of onboarding and the reduced focus on how to read a chart lead us to this contribution. Thus, we wanted to explore different aspects of visualization onboarding: general effect of visualization onboarding on user performance (study 1), the effect on user performance of different types of visualization onboarding (study 2), and the use of in-situ visualization onboarding in the VA tool Netflower (study 3).

## 7.1. Lessons Learned Study 1 & 2

*Lessons Learned Study 1:* Based on the qualitative feedback of the onboarding concept (study 1), onboarding was described to be *easy to use* and *easy to understand*. Moreover, the *easy to understand example (monthly changes in weather)* supported the visualization literacy for *new kind of graphs*. Also the results of the quantitative measures showed that the participants ranked the onboarding concepts as *easy to use*, *easy to comprehend*, and *easy to follow* with regard to the design. Also, participants would be willing to use the onboarding concept when working with new visualization types.

In fact, onboarding elevates the improvement of response accuracy. However, assessing the effects on visualization type level, we could not observe differences in improvements between surveys with and without onboarding. On the contrary, the results indicate that the highly familiar visualization type — bar chart — does not require onboarding and results in faster answers after multiple question iterations (pre- and post-tasks). Here, we can refer to learning effects based on repetition. Items that are practiced a lot acquire a high activation in our memory and are retrieved faster [12]. Feedback on the widely known visualization type—the bar chart—indicates to be "self-explanatory" or "straightforward". This finding also coincides with a low interaction time and hardly any mouse movements with the onboarding. The horizon graph for instance showed, in spite of an intermediate level of familiarity, both, a high improvement in response accuracy as well as response time. Further, participants of the horizon graph and the change matrix emphasized "the organization of the onboarding" and the "choice of colors" particularly positively. Similarly, their interaction time with the onboarding concept increased. Similar to the improvements, the interaction time with the onboarding concepts increased gradually with the level of familiarity. Thus, the higher the experience with a visualization, the lower the interaction time with the onboarding. Interestingly, for the most unfamiliar visualization type—the parallel coordinates plot—the feedback on onboarding diverged. On the one hand, participants appreciated that the "important words have been marked so that you could more easily find out, what the step was about" which goes along with our consideration of basing visualization literacy descriptions on the Vega-Lite specification and highlight respective visual properties. On the other hand, some participants where overwhelmed by the amount of in-formation displayed due to a lack of literacy. To reduce this information overload, features should be "introduced slowly and not at the same time". This leads to a gradual approximation of new functionality with highly unfamiliar and complex visualization types.

One participant mentioned the following: "Unfortunately, I still don't think I have a handle on how to understand and read these charts. I still have questions. I need to be able to redo the questions I was first asked to see if I can get them right. I need to not only read the step by step guide, but put what I have learned into practice to make sure I truly understand what I learned." This approach appears to be highly interesting as integrating challenges and small tasks in the onboarding concept can provide a format to check if the given text was understood correctly.

*Lessons Learned Study 2:* Reflecting on the feedback from study 2, we summarize the most important insights. We realized that independently of the visualization type and method applied, an easy-to-understand data set and concrete examples on how to read the chart, support and increase the comprehension are vital. Furthermore, qualitative feedback indicated that the to-the-point descriptions make it easier to absorb information. Based on the comments, the competence to learn new interaction techniques can be increased by video tutorials. Likewise, the understanding of introductions is enabled by interactive and linked descriptions of the visualizations B and the described steps D (as used for the step-by-step guide and the scrollytelling tutorial). Finally, it is important to consider which visualization types (predominantly unknown and new ones) require an introduction in order to support the user when needed.

## 7.2. Lessons Learned In-situ Evaluation – Study 3

In this investigation, we distinguished between performance measures and attitudes of the participants. In general, we found that there were no significant differences in the performance of participants using the tutorial and scrollytelling. This may be due to the fact that the visualization used was not that demanding for computer science students. Many participants mentioned that they figured out how the features of the visualization worked themselves. Nevertheless, there were single tasks that were difficult for the participants. Even in these cases, participants sometimes did not use the onboarding systems to solve these tasks. Integrating solved tasks (worked examples) and small challenges might help to overcome this problem.

On the other hand, the attitude of the participants was decidedly in favor of scrollytelling. Participants reported that they used scrollytelling more often than the tutorial before starting to work. Many more of them stated that they found scrollytelling more appealing than the tutorial. Features they found especially appealing were the fact that single components of the visualization appeared gradually and that the onboarding system was integrated into the system and did not interrupt the workflow. Positive features of the tutorial were that it provided a nice overview and contained videos. The attitude concerning videos is mixed, there were some participants who explicitly stated that they did not like videos.

*Summary of Lessons Learned from all Studies* Taking together all the results of the three studies we can formulate the following lessons learned towards visualization onboarding:

- Subjective experiences levels per visualization type do not correlate with a high answer correctness.

- Participants prefer the video tutorial to learn how to interact with the parallel coordinates plot (over scrollytelling and step-by-step guide).

- In-situ scrollytelling is highly preferred over a tutorial due to its integration in the Netflower tool and gradual guidance through the instructions.

## 7.3. Towards Design Guidelines for Visualization Onboarding Methods

To facilitate the creation of visualization onboarding methods, we summarize some core advise we have learned during our research. In the following, we list the design guidelines.

- In general, onboarding systems that are **integrated** into the system are appreciated. Users prefer systems that do not disrupt the workflow. In such systems, important **features** should be **highlighted**.

- Use an **easy-to-understand data set** and **concrete examples** on how to read the charts, support and increase comprehension are vital.

- **To-the-point descriptions** make it easier to absorb information (step-by-step).

- Use **videos** or small animations as onboarding method to explain interaction techniques. In contrast, videos are appreciated by a minority. Other users have a negative attitude. It depends on the target group whether they can be successfully deployed.

- Some users tend to ignore onboarding systems, even if they struggle. These users have to be **motivated to use onboarding**.

## 7.4. Limitations

**Study Design.** A limitation of our conducted study 1 and 2 is the study design. We designed the study, as illustrated in Figure 5, integrating pre- and post-questions to assess the improvement of participants. Participants assigned to the with or without onboarding condition had to answer those questions. We couldn't find any statistical significance for answer correctness between the groups. We assume that the participants of the without onboarding condition could increase their answer correctness because of the repetitive nature of the questions. So there was already a learning effect visible for participants who did not interact with the visualization onboarding.

**Experience Level and Answer Correctness.** Besides the limitation of the study design, we can report on lessons learned regarding the answer correctness (in %) in general and the answer correctness with respect to the subjectively-rated level of experience of MTurk workers. In line with the previous work by [10, 9, 24], we see also poor answer correctness over different onboarding concepts (cf. Study 2 3.1 and Table 2). Our results show answer correctness around 50%. For the Baseline (no onboarding) we can report an answer correctness of the tasks of 52%. This is an alerting results and indicates the need for visualization onboarding. When comparing the subjective assessment of the experience with different visualization techniques to the answer correctness of pre-questions, the analysis revealed discrepancies. When assessing an influence of the subjectively rated experience levels with the visualization types (pre-questions), we observed that highly experiences participants did not show a high answer correctness. In contrast, rather inexperienced users achieved higher performance measures.

A similar phenomenon occured with the computer science students in study 3. They were confident about their ability to solve tasks with the Netflower tool. Nevertheless, in certain task categories requiring less obvious modes of interaction, the results were fairly mixed. This indicates that users should consult the onboarding system more often than they really do. In future work, we plan to address this issue, that is, how to motivate users to work with the onboarding system more intensively and check their solutions more often. Possible solutions for this might be to add tasks with solutions (worked examples) or small challenges to the onboarding system, so that users can check whether their method of problem solving was appropriate.

**Target Group.** Additionally, the analysis of the interaction time with the visualization onboarding (see Study 1) showed that the MTurk workers may not take enough time with the onboarding concept itself to carefully read through the text, for example. This is a limitation of our study as we should have applied a scaling incentive system were participants get paid more for correct answers to encourage them engaging more with the onboarding concept.

Participants who used MTurk worked fairly briefly and got a visualization with reduced interaction possibilities. The computer science students in study 3 worked for a longer time and used a fairly complex tool with many different possibilities to solve tasks. This difference might be responsible for differences in attitude to onboarding systems in study 1 and 2 and in study 3. The scrollytelling method is probably more useful for complex visualization tools when users need more detailed information about the tool and flexible access. All three studies indicate that an integrated, step-by-step approach was appreciated more by users than a stand-alone onboarding system. The integrated system helps users to relate the information of the onboarding system to the features of the visualization, and the the step-by-step procedure avoids information overload of inexperienced users. Especially in study 3, we found that participants sometimes had difficulties to know when they needed help. They were overconfident about their ability to solve problems and primarily relied on the possibility to find out about the features of the visualization by trial and error. Therefore, they had specific difficulties to solve tasks that required them to use features of the system in a way that was not obvious. Future work should clarify how this problem could be solved. There is some indication that tasks with solutions and small challenges can help to overcome this problem.

# 8. Conclusion and Future Work

In this paper, we present the results of three studies with MTurk workers as well as students (see Figure 1) to investigate how visualization onboarding affects user performance at different levels. Firstly, we assessed the effect of visualization onboarding in performance measure between different visualization types —for a bar chart, horizon graph, change matrix, and parallel coordinates plot (study 1). We identified significant differences in the answer correctness between the four visualization types. Horizon graph and bar charts show more correctly answers than the other visualization types. Relating the response time, the fastest answers where given to the bar chart, horizon graph, the change matrix and then the parallel coordinates plot. Over all surveys including all four visualization types, we could not determine statistically significant differences between the answer correctness in surveys with or without onboarding. The participants in the control condition showed statistically significant faster answers than participants in the onboarding condition. Thus, based on all visualization types, we could not identify differences or improvements. Furthermore, we also assessed differences between task types of *reading the data*, *reading between the data*, and *reading beyond the data*. Irrespective of the onboarding method and the visualization type, *reading the data* questions showed the highest mean rank, followed by *reading between the data*, and *reading beyond the data* question types. Overall, the onboarding has influences answer correctness for the change matrix and the results for the most unfamiliar visualization type — the parallel coordinates plot — shows that onboarding is needed to improve the performance while interacting with the chart.

We conducted a further study with MTurk workers to investigate the effect on user performance for four different onboarding types (step-by-step, scrollytelling, video tutorial, and in-situ scrollytelling in Netflower) in study 2 and 3. The results of study 2 showed that no statistically significant difference in the answer correctness for the post-questions could be determined between the different onboarding types. However, the response time showed that faster answers were given after a scrollytelling tutorial, followed by the step-by-step guide, the condition without any onboarding, and the video tutorial. The analysis of onboarding and task difficulty by Friel et al. [23] showed that participants *reading the data* tasks were answered the fastest, followed by *reading between the data*, and *reading beyond data*. The analysis of the qualitative feedback revealed that independently of the visualization type and method applied, an easy-to-understand data set and concrete examples on how to read the chart, support and increase the comprehension are vital.

For study 3 we gathered data toward user experience with using the in-situ scrollytelling for the VA tool Netflower. The results of the evaluation with students showed that while there was no significant difference in answer correctness (see 6.6), we did find a difference in attitude: Participants reported to prefer in-situ onboarding to an external one.

We plan to further explore how to integrate such onboarding concepts in VA systems supporting novice users in understanding the visual encoding and interaction concepts. Additionally, we plan to work on concepts to provide an appropriate way of semi-automatically implementing onboarding concepts by using declarative language.

*Acknowledgements* The authors wish to thank Victor Adriel de Jesus Oliveira for his valuable feedback. This work was funded by the Austrian Ministry for Transport, Innovation and Technology (BMVIT) under the ICT of the Future program via the SEVA project (no. 874018), as well as the FFG, Contract No. 881844: "Pro²Future is funded within the Austrian COMET Program Competence Centers for Excellent Technologies under the auspices of the Austrian Federal Ministry for Climate Action, Environment, Energy, Mobility, Innovation and Technology, the Austrian Federal Ministry for Digital and Economic Affairs and of the Provinces of Upper Austria and Styria. COMET is managed by the Austrian Research Promotion Agency FFG."